\documentclass[prl,aps,superscriptaddress,a4paper,twocolumn]{revtex4}
\usepackage[dvips]{graphicx}
\usepackage[utf8]{inputenc}
\usepackage[T1]{fontenc}
\usepackage{float}\usepackage[dvipsnames]{xcolor}

\usepackage{epsfig,amsmath,amsfonts,amssymb,
color,
}
\usepackage{siunitx}
\usepackage{comment}
\begin{document}

\title{Kinetic theory and shear viscosity of
dense dipolar hard sphere liquids 
}

\author{Faezeh Pousaneh$^{\scriptscriptstyle{}}$\footnote{Electronic address: \texttt{faezeh.pousaneh@ntnu.no}.}}  
\affiliation{Department of Mechanical and Industrial Engineering, Norwegian University of Science and Technology, 7491 Trondheim, Norway}
%
%
\author{Astrid S. de Wijn$^{\scriptscriptstyle{}}$\footnote{Electronic address: \texttt{astrid.dewijn@ntnu.no}.}}
\affiliation{Department of Mechanical and Industrial Engineering, Norwegian University of Science and Technology, 7491 Trondheim, Norway}
\affiliation{Department of Physics, Stockholm University, Sweden}

\begin{abstract}
Transport properties of dense fluids are fundamentally challenging, because the powerful approaches of equilibrium statistical physics cannot be applied.
Polar fluids compound this problem, because the long-range interactions preclude the use of a simple effect-diameter approach based solely on hard spheres.
Here, we develop a kinetic theory for dipolar hard-sphere fluids that is valid up to high density.
We derive a mathematical approximation for the radial distribution function at contact directly from the equation of state, and use it to obtain the shear viscosity.
We also perform molecular-dynamics simulations of this system and extract the shear viscosity numerically.
The theoretical results compare favorably to the simulations.
\end{abstract}
\maketitle

Transport properties of dense fluids are fundamentally challenging, because it is a many-body problem out of equilibrium.
A statistical approach is needed, but the powerful approaches of equilibrium statistical physics cannot be applied.
Current theoretical approaches to transport in dense fluids are based on hard spheres and Enskog's heuristic extension of the Boltzmann equation and kinetic theory of gasses and liquids~\cite{Viswanath:07:0}.
The only alternative to this is to resort to purely computational methods (see, for example~\cite{stephan:79:0,Ashurst:75:00,Daschakraborty:12:00,Theiss:19:00}).
Kinetic theory was heavily developed in the 60s and 70s, but little progress has been made since.  In particular, there  is no analytical description of high-density fluids consisting of anything more complicated than simple hard spheres (HS).
This is a fundamental limitation in our current understanding, but also particularly problematic in practical applications, where kinetic theory is widely used in combination with empirical information and effective diameters to predict viscosities of some non-polar complex liquids~\cite{Astrid:12:00,vesovic:12:00}.

Here, we develop kinetic theory of polar fluids, especially focusing on the viscosity.
Polar fluids are a textbook example of systems where the hard-sphere approach fails, because the long-range electrostatic interactions are captured badly by instantaneous collisions.
They are also ubiquitous in nature, for example in the form of water, and are increasingly important in applications in biotechnology and other fields.
The physics of these systems  cannot be described by simple hard spheres with an effective diameter.
Moreover, Molecular Dynamics (MD) simulations involving electrostatic interactions are extremely computationally demanding and anyway cannot provide the fundamental understanding that is needed.

We choose to focus on the shear viscosity, as it is one of the most important transport properties of a fluid for practical applications.
It plays a crucial role in for example lubrication and pipe flow.
The viscosity of polar fluids is receiving increasing interest in practical applications, for example as the basis of environmentally-friendly lubricants, and are very promising for low-friction applications, as demonstrated for instance by the amazing effectiveness with which water-based synovial fluid lubricates our joints~\cite{Bayer:08:00,Das:03:00}. 

We derive an analytical kinetic theory for the viscosity of a simple model for a polar fluid, dipolar hard spheres (DHS). 
Our theoretical approach is based around Enskog's extension of the Boltzmann equation to high densities (BEk).
In order to incorporate the soft and long-range electrostatic interactions between the dipoles, we extend this theory, which is originally based on simple shapes and simple interactions especially HS at low densities.
We do this by explicitly including the dipole-dipole interaction into the Radial Distribution Function (RDF).
We calculate the RDF using the method  of~\cite{Madden:78:00,Lee:89:0,Kusalik:88:0,patey:85:00}
from the Helmholtz free energy of DHS derived by Elfimova {\it {et al.}}~\cite{Elfimova:16:00}.
In order to verify our theoretical results, we compare them to MD simulation of dipolar pseudo hard spheres.
Our result can be used in a straight-forward manner to also calculate other transport properties such as thermal conductivity and diffusion coefficient.

BEk theory centers around the Boltzmann equation and deals with collision probabilities and collision dynamics.
Boltzmann's original equation contains a crucial low-density approximation: the Sto\ss{}zahlansatz, which states that when particles collide they are uncorrelated.
Solving the Boltzmann equation for transport coefficients is nontrivial, but general solutions were derived by Chapman and Enskog~\cite{Chapman:52:0}.
The general form of  the zero-density viscosity is found to be
\begin{equation}
\eta_0=\frac{5}{16\sigma^2 \Omega^{*(2,2)}}\sqrt{\frac{mk_B T}{\pi}}~,
\label{eq:eta0}
\end{equation} 
where ${\Omega^{*(2,2)}}$  is the collision integral  which  depends on the interactions.
For HS, ${\Omega^{*(2,2)}}=1$.
With considerable effort, zero-density viscosities can also be derived for slightly more complicated interaction models, such as rough spheres~\cite{Chapman:52:0}, spherocylinders~\cite{spherocylinders}, and hard spheres with embedded point dipoles (DHS)~\cite{dhs-low-density}.

At higher density, the equations for collision rates and dynamics become more complex, in general including correlated collisions.
Enskog devised a heuristic way to incorporate some correlated collisions at higher densities, but this approach is currently limited to HS~\cite{Chapman:52:0} and chains of hard spheres~\cite{Astrid:08:00}, but not other types of interactions.
Enskog's approach produces good agreements with simulations of HSs and experiments of very simple fluids only for low- to mid-density ranges and fails at high densities, since it still does not take into account correlated collisions.
Nevertheless, Enskog's theory, though still approximate in nature, has provided a useful theoretical basis for both understanding and predicting the transport properties of hydrocarbons with short-range interactions only, including some molecules with much more complex geometry than HS~\cite{KUMARI:09:0,Castillo:93:00,Astrid:08:00}.

In order to obtain theoretical results for the  viscosity of  DHS, we start from the  Enskog's theory for a simple dense fluid.
Enskog's expression for the viscosity is~\cite{Chapman:52:0,Santos:16:0,Viswanath:07:0,Sengers:00:0,Lucas:79:0}
\begin{equation}
\eta= \eta_0 \bigg [ g(\xi)^{-1}+0.8 V_\mathrm{excl} \rho + 0.776~ V^2_\mathrm{excl} \rho^2 g(\xi) \bigg]~,
\label{eq:ENSHS}
\end{equation} 
where $V_\mathrm{excl} $ is the excluded volume of HS,  $V_\mathrm{excl}=(2\pi/3) \sigma^3 $, $\xi= \pi \sigma^3 \rho/6$ is the volume fraction,  and $g(\xi)$ is the RDF at contact.
The RDF in is the spherical component of the pair-distrition function. 
There are a number of ways to obtain good approximations for the RDF at contact of HS, such as from the  Carnahan-Starling equation~\cite{Carnahan:69:00}, which gives
\begin{equation}
g_\mathrm{HS}(\xi) = \frac{ 1 - \frac12 \xi}{(1-\xi)^3}~.\label{eq:CSRDF}
\end{equation}
The zero-density limit for viscosity [see Eq.~\ref{eq:eta0}] for some polar interactions have been obtained.
In Ref.~\cite{monchick:61:0}, the collision integral for the zero-density viscosity of polar gas was calculated for the Stockmayer potential.
Chung {\it {et al.}} developed an empirical formula which works well for the viscosity of real dilute gasses~\citep{chung:84:00}.

Our approach for high densities is to develop the radial distribution function of DHS and apply it to the Enskog expression.
The interaction between two DHSs $i$ and $j$ with diameter $\sigma$ and dipole moments $\mu$ at distance $r$ is given by a sum of hard sphere ($U_{ij}^{HS}$) and dipolar ($U_{ij}^{D}$ ) terms:
\begin{eqnarray}
U_{ij}^{HS}= \begin{cases} 
\infty; \hspace{1.77cm}  r<    \sigma \\
  0  ; \hspace{2cm}     r \geq   \sigma
  \end{cases}
  ~,\\
U_{ij}^D= \bigg[\frac{  \boldsymbol \mu_i \cdot \boldsymbol \mu_j }{r_{ij}^3}  - \frac{3( \boldsymbol\mu_i \cdot  \boldsymbol r_{ij})( \boldsymbol\mu_j \cdot  \boldsymbol r_{ij})}{r_{ij}^5} \bigg]~,
\end{eqnarray} 
with the dipolar coupling constant $\lambda= {\mu^2}/({k_B T  4\pi \epsilon_0 \sigma^3})$.

In recent years, considerable effort has been made on development of the theoretical expression for the equilibrium properties of DHS \cite{Ivanov:07:00,Elfimova:10:1,Elfimova:10:0,Elfimova:12:00,Elfimova:16:00,Ladanyi:99:00,Henderson:11:0}. 
The Helmholtz free energy of DHS can be written relative to that for a regular HS fluid $F^\mathrm{HS}$ as
\begin{equation}
 F^\mathrm{DHS} =F^\mathrm{HS} + F^\mathrm{D},
\end{equation}
where  $ F^\mathrm{D}$ is the excess free energy due to the electrostatic interaction between the dipoles.
The most common approaches for dealing with DHS is thermodynamic perturbation theory with a 
Pade approximation and mean spherical approximation (MSA).
However, because these are lower-order theories with respect to   $\lambda$
they do not give accurate results for low densities and  virial coefficients~\cite{Ladanyi:99:00,Elfimova:12:00}.

In order to get around this problem, Elfimova {\it et al.}~\cite{Elfimova:12:00} introduced a logarithmic representation of the free energy.  The result converges faster, since the logarithm of a polynomial is less sensitive to the truncation of the polynomial. 
The excess free energy is then written as~\cite{Elfimova:16:00}
\begin{equation}
\frac{\beta F^\mathrm{D}}{N} =-\ln \bigg [1+  \sum_{n=1}^{\infty} n^{-1} I_{n} \xi^n\bigg ].
\end{equation}
The coefficients $I_n$ are obtained from the regular virial coefficients for DHS.
Elfimova {\it et al.}~\cite{Elfimova:16:00} keep up to the fifth virial coefficient, corresponding to $n=4$ and give explicit expressions for $I_{1,2,3,4}$.
This theory accurately captures the free energy and compares favorably with computer simulation for $\lambda \le 4$, even at high value of the particle volume fraction $\xi \le 0.5$.

We obtain the RDF from the DHS free energy using the equation of state (EOS) \cite{Lee:89:0,Pippo:77:00},
\begin{equation}
\frac{PV}{Nk_BT} =1+ \frac{\langle U_\mathrm{pot} \rangle }{Nk_BT} + \frac{2\pi \rho}{3} \sigma^3 g(\xi)~,
\label{kusalik}
\end{equation}
where  $\langle U_\mathrm{pot} \rangle$ is the interaction potential.
We apply the  thermodynamic relations
$P=-\frac{\partial  F}{\partial V} \vert_{N,T}$ 
and
$\langle  U_\mathrm{int}\rangle=\frac{\partial (\beta F)}{\partial\beta}$
to obtain the pressure and internal energy. The interaction potential is then obtained as
\begin{equation}
\langle  U_\mathrm{pot}\rangle=-\frac{J_1(\lambda)\xi+J_2(\lambda)\xi^2+J_3(\lambda)\xi^3+J_4(\lambda)\xi^4}{1+I_1(\lambda)\xi+\frac{1}{2}I_2(\lambda)\xi^2+\frac{1}{3}I_3(\lambda)\xi^3+\frac{1}{4}I_4(\lambda)\xi^4}  ,
\end{equation}
where $J_i(\lambda)= \frac{\lambda}{i}\frac{\partial I_i(\lambda) }{\partial\lambda} $.
Finally, we find the RDF at contact
\begin{eqnarray}
\lefteqn{g(\xi) =
    \frac{1}{4 \xi}
    \bigg[Z^{HS}-1} &\nonumber\\
    & +\frac{L_1(\lambda)\xi+L_2(\lambda)\xi^2+L_3(\lambda)\xi^3+L_4(\lambda)\xi^4}{1+I_1(\lambda)\xi+\frac{1}{2}I_2(\lambda)\xi^2+\frac{1}{3}I_3(\lambda)\xi^3+\frac{1}{4}I_4(\lambda)\xi^4}   \bigg] ,
\label{eq:g}
\end{eqnarray}
where 
\begin{equation}
L_i(\lambda)= J_i(\lambda) - I_i(\lambda).
\end{equation}

The viscosity is then obtained by substituting this into Eq.~(\ref{eq:ENSHS}).
We compare our  theoretical results to MD simulations.
The integration algorithms typically used for MD depend on smooth interaction, and cannot be applied to instantaneous collisions.  This is worsened by the presence of long-range electrostatic interactions, which require additional techniques that also depend on smooth force fields.
We circumvent this issue by employing a pseudo hard sphere model (PHS) introduced by Jover {\it {et al.}}\cite{jover:12:0}.
The PHS potential is of Mie form where the typical powers of the LJ potential 12/6 are replaced by 50/49:
\begin{equation}
U^{(50,49)}(r)=
  \begin{cases} 
    50 (\frac{50}{49})^{49}
    \epsilon \big[ \big(\frac{\sigma}{r}\big)^{50 }-\big(\frac{\sigma}{r}\big)^{49 }  \big]+\epsilon & r<   \frac{50}{49} \sigma \\
   0 & r \geq   \frac{50}{49} \sigma
   \label{eqPHS}
  \end{cases}~.
\end{equation}
Jover {\it {et al.}} verified that this potential accurately captures the thermodynamics, structures, and dynamics of the HS system.
It produces good results at reduced temperature $T^*= \frac{\epsilon}{k_B T} =2/3$. 
This model has been shown to accurately describe the fluid-solid equilibrium~\cite{vega:13:0} as well as the viscosity of HS~\cite{faezeh:19:00}.

\begin{figure}
\includegraphics[scale=0.5]{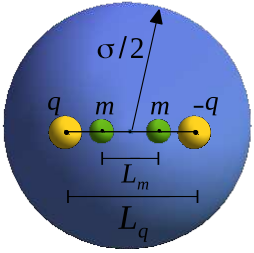}
\caption{Schematic representation of the DHS model. The central atom with diameter $\sigma$ is connected to two dummy massive atoms with mass $m$ and two oppositely charged virtual sites with charges $\pm q$ each at distance $L_q$.}
\label{DHS}\vspace{0.1cm}
\end{figure}

We use Gromacs version 5 to integrate the equations of motion and the PHS potential is implemented as a tabular form as in Ref.~\cite{faezeh:19:00}.
Our DHS consist of 5 particles on a line, as  shown in Fig.~\ref{DHS}.  The central particle has no charge or mass, but interacts with the central particles of the other DHS through a PHS potential.
Two massless particles of opposite charges $q$ and $-q$ are at equal distance $L_q/2$ from the center on either side and give rise to the dipole.
There are also two dummy mass particles of mass $m$ on each side at distance $L_m/2$, controlling the moment of inertia.
We use $L_q/\sigma=0.224$, since the point dipole model has been found to agree well with the extended dipole model up to  $L_q/\sigma=0.3$~\cite{Theiss:19:00,Drunsel:14:00,Ballenegger:04:00}. In our model $L_m/\sigma=0.2$.

We simulate this system for different dipole moments corresponding to $\lambda=1,2,3,4$.
 Our simulation box contains $N=1000$  DHS particles and $N=6000$ for dilute cases, $\rho\sigma<0.15$. All simulations have been carried out at reduced temperature $T^* =\epsilon/k_BT=2/3$.
For each systems with different  $\lambda$ we perform simulations for a range of densities $\rho^*$ between $0$ and $1$.
In what follows, all units are dimensionless as: $t^*=t[ {k_B T}/({\sigma^2 m})]^\frac12$, $r^*= \frac{r}{\sigma}$, $\rho^*= \rho \sigma^3=\xi 6/\pi$ and $P^*=  P \sigma^3/(k_B T)$,  $\mu^*=  \mu (k_B T\sigma^3 4\pi \epsilon_0)^{-\frac12}$, $\lambda= {\mu^*}^2$, $\eta^*=\eta \sigma ^2 /( mk_B T)^\frac12 $, where $\rho$ and $P$  denote number density and  pressure respectively and $\eta$ is viscosity.  The reduced volume fraction is $ \xi^*=\frac{\pi \rho^* }{6}$.
The electrostatic interactions are treated using the Particle Mesh Ewald (PME) method with cut-off length of $2.6 \sigma^*$.
Time steps for simulations is $\delta t^*=0.0011$.

We first equilibrate the system and verify the equation of state (EOS), before moving on to the RDF and viscosity.
Equilibration was performed in the NVT ensemble with the velocity-rescale thermostat for   $t^*= 10^5$  to $t^*=6 \cdot 10^5$ depending on the system.

\begin{figure}[]
\parbox{43mm}{(a)\hfill\strut}\parbox{43mm}{(b)\hfill\strut}\\[-2.5ex]
\includegraphics[width=43mm]{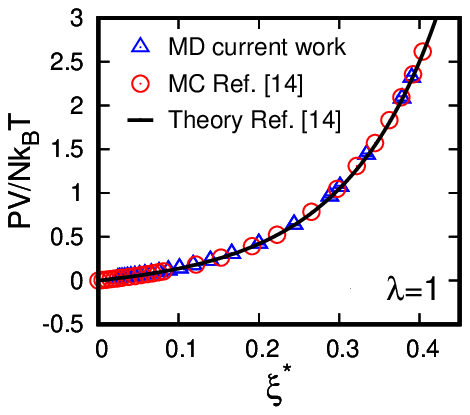}\includegraphics[width=43mm]{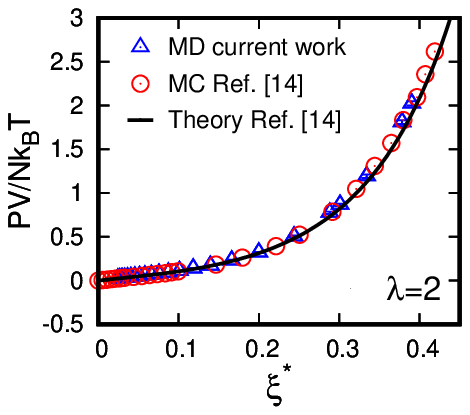}\\[-1ex]
\parbox{43mm}{(c)\hfill\strut}\parbox{43mm}{(d)\hfill\strut}\\[-2.5ex]
\includegraphics[width=43mm]{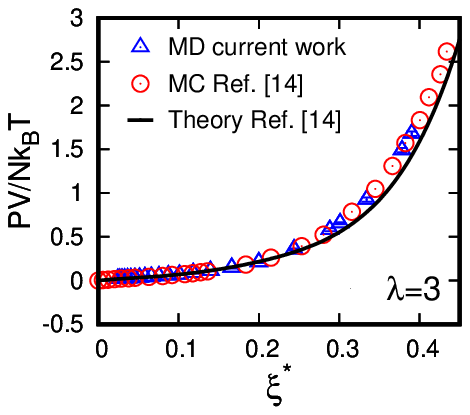}\includegraphics[width=43mm]{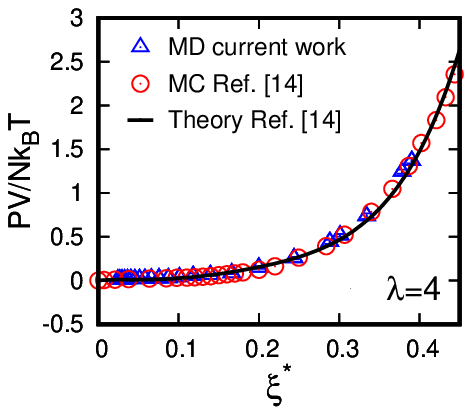}\\[-2ex]
\caption{The equation of state of DHS fluids from current simulations for $\lambda=1,2,3,4$ (triangle data) and from monte carlo simulations by Elfimova {\it {et al.}}~\cite{Elfimova:16:00} (circle data). Solid lines are the theoretical expression in Ref.~\cite{Elfimova:16:00}.}
\label{com}\vspace{0.cm}
\end{figure}

Fig.~\ref{com} shows  the  EOS for DHS obtained from current simulations (blue triangle data), from previous Monte-Carlo simulations~\cite{Elfimova:16:00} (red circle data) and the theoretical expression of EOS in Ref.~\cite{Elfimova:16:00}.
Our MD simulations correspond well to both.

\begin{figure}[]
\vskip1.5ex
\parbox{43mm}{(a)\hfill\strut}\parbox{43mm}{(b)\hfill\strut}\\[-5.0ex]
\includegraphics[width=43mm]{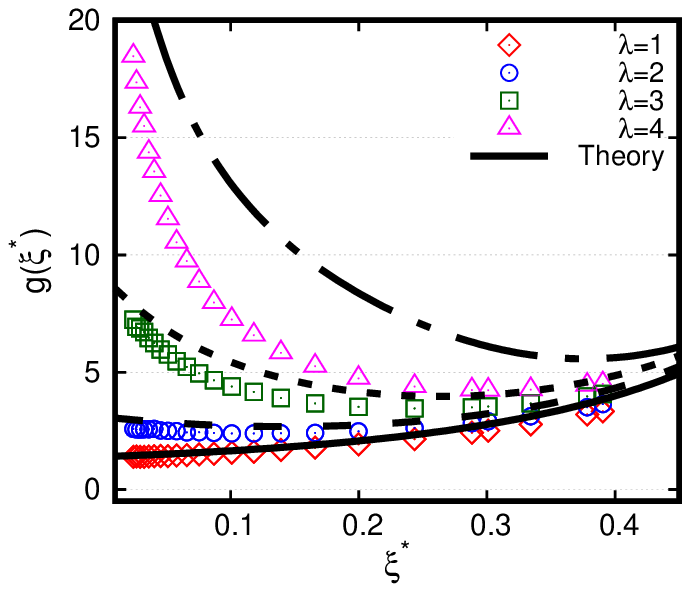}\includegraphics[width=43mm]{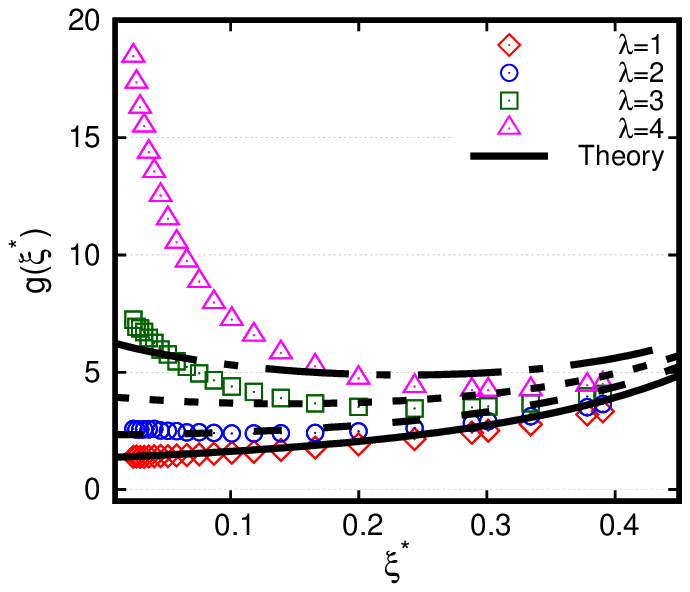}\\[-2ex]
\caption{Rdf at contact from simulations (data) along  with (a) the RDF at contact from the present work, Eq.~(\ref{eq:g}),  and (b) the RDF obtained from the Pade approximation \cite{Rushbrooke:73:00}. In both plots $\sigma=1$.}
\label{RDF_plots}
\end{figure}

After equilibration, we run the simulations in the NVT ensemble for an additional interval of $t^*=1000$
and obtain the full RDF for each density and $\lambda$.
The RDF at contact is given by the maximum values of RDF.
Simulation results of RDF at contact are shown in Fig.~\ref{RDF_plots}(a)  along with our theoretical expression Eq.~(\ref{eq:g}), both with  $\sigma=1$.
For comparison, Fig.~\ref{RDF_plots}(b) shows the results obtained by Rushbrooke {\it {et al.}}\ using the Pade approximation~\cite{Rushbrooke:73:00,jog:99:00}.
The new theory developed here describes the simulation results significantly better and captures the trends relative to the Carnahan-Starling results.

  We continue to run the system in the NVT ensemble for $t^*= 45000$ up to $t^*= 90000$ depending on the system (for more dilute ones longer time is needed to get enough collisions).
To minimize the influence of the thermostat, the temperature is controlled using a Berendsen thermostat with a slow coupling with a characteristic time of  $t^*=11$.
We obtain the shear viscosity of the DHS model using the transverse-current auto-correlation function (TCAF) method~\cite{palmer:94:0}.
More details on our use of this method can be found in~\cite{faezeh:19:00}.

\begin{figure}
\vspace{0.5cm}
\includegraphics[width=6cm]{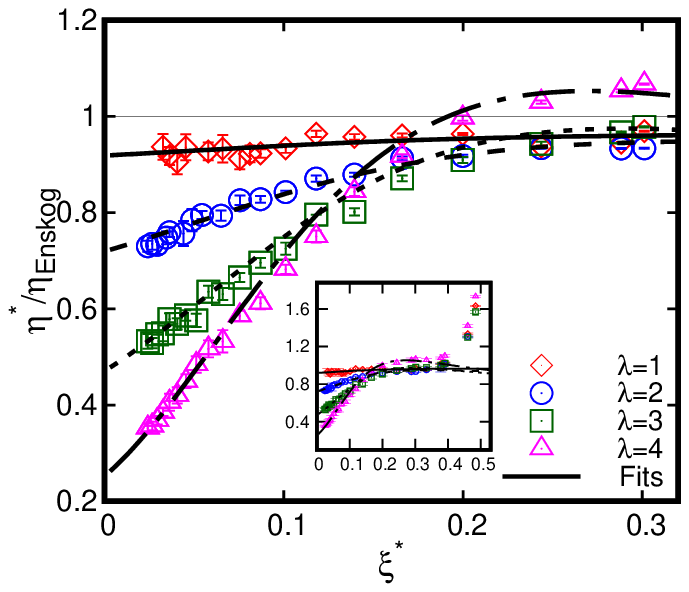}\\[-5cm]
~~(a)\hfill\strut\\[-2ex]\vskip6cm\vskip-2ex
\vspace{-0.5cm}
\includegraphics[width=6cm]{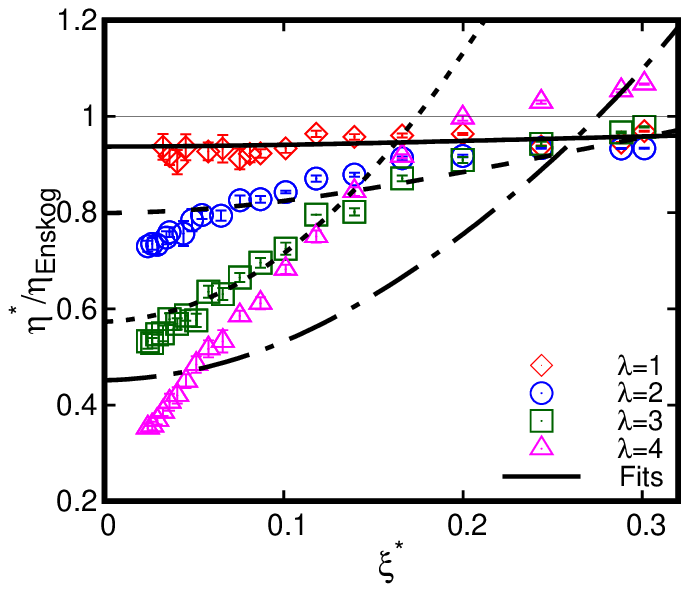}\\[-5cm]
~~(b)\hfill\strut\\[-2ex]\vskip6cm\vskip-2ex
\vspace{-1cm}\caption{Shear viscosity of DHS  from simulations (data points) fitted to (a) the theory developed in the present work and (b) the Enskog theory of HS with an effective diameter. The inset in (a) shows the same data in different scale to include the higher densities.  }
\label{DHS_eta} 
\end{figure}

\begin{figure}[]
\hspace{-3.5cm}
\includegraphics[scale=0.5]{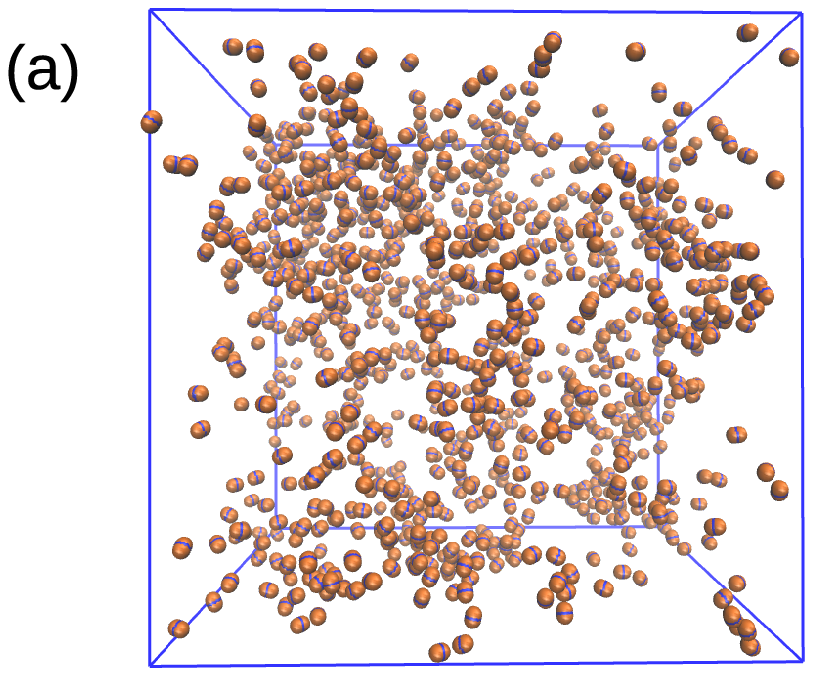}
\hspace{-2.7cm}
\includegraphics[scale=0.55]{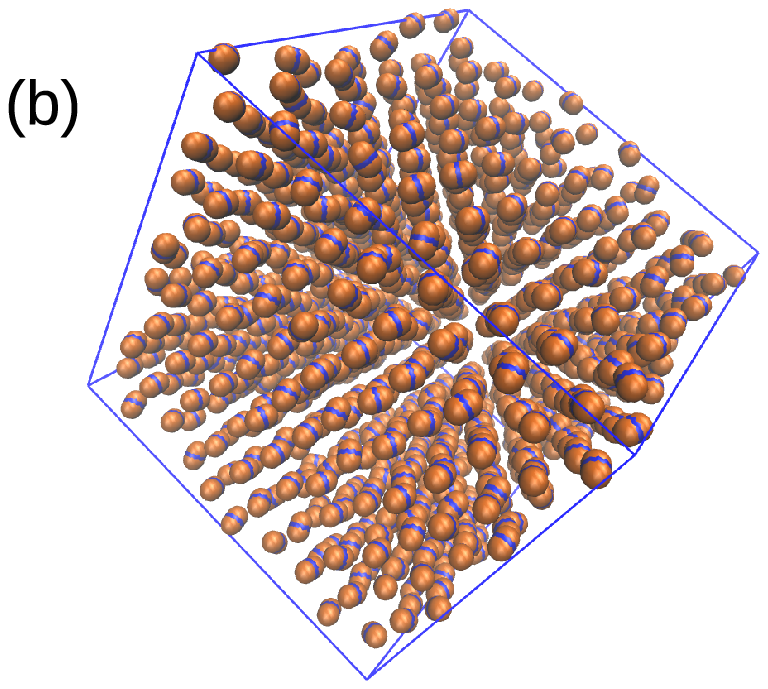}
\hspace{-4.cm}
\caption{Snapshots of the simulation results for system with  $\lambda=4$ for two densities $\xi^*=0.03$  (a) and $\xi^*=0.49$ (b). At low densities we observe clustering of the dipoles in an otherwise disordered fluid, while at high densities there is more orientational structure.  These behaviors are a result of the strong directionality and long-range interactions and have an impact on the viscosity.}
\label{snapshot}
\end{figure}

The shear viscosities obtained from the simulations are shown  in Fig.~\ref{DHS_eta}(a) for different $\lambda$. According to the data, the relative shear viscosity decreases upon increasing the dipole moment, whereas it shows an opposite behavior for higher densities. This is because at lower densities the dipolar particles form chain-like structure which decreases the collision rates, and consequently the viscosity.  At high densities strong dipole moments cause the system to
form ordered structures, which have a higher viscosity (up to infinity) than
a noninteracting disordered fluid.  In addition, Enskog theory for hard spheres is known to break down at higher densities even in fluids.  To visualize the reason for why the viscosities deviate from the Enskog
expressions for hard spheres without dipoles, we show examples of snapshots
from simulations in Fig.~\ref{snapshot}. The snapshots are for system with $\lambda=4$  and two
different densities $\xi^*=0.03$  and $\xi^*=0.49$.  Similar structures are reported
by simulations for dipolar fluids (and ferromagnetic particles) \cite{Weis:93:00,Weis:03:00,Weis:06:00,Camp:00:00,wei:11:00} and
also by experiments \cite{Klokkenburg:07:00,Klokkenburg:06:00,Butter:03:00,Butter:03:01}.  

We compare the simulation results to the theoretical results, which are given by Eq.~(\ref{eq:g}) combined with~(\ref{eq:ENSHS}) and~(\ref{eq:eta0}).
Since we do not have the exact low-density limit for $\eta_0$ for our system, we introduce ${\Omega^{*(2,2)}}$ as a fit parameter. 
We  estimate the range of sensible values from the results of Ref.~\cite{monchick:61:0} for the Stockmayer potential with a point dipole, to be in the range of $1$ to $3$ corresponding to $\lambda$ up to $\lambda=5$.
We use an effective dipole moment $\mu_\mathrm{e}$ as a fit parameter, rather than the hard-sphere diameter.
The fit parameters ${\Omega^{*(2,2)}}$ and $\mu_\mathrm{e}/\mu$ are given in Table.~\ref{fit_eta}.
The obtained values  of ${\Omega^{*(2,2)}}$ indicates collision integrals increase by increasing the dipole moments in agreement with the trends found for the zero-density viscosity of the Stockmayer potential~\citep{monchick:61:0} and values in the expected range.

\begin{table}
\parbox{42mm}{
fit parameters for DHS
\begin{tabular}{|c |c|  c|}
 \hline
&  \hskip0cm  ${\Omega^*(2,2)}$  \hskip3cm &  $\mu/\mu_\mathrm{e}$ \\
      \hline
 $\lambda=1$  &   \hskip0.2cm 1.05 &   1.78\\
      \hline
        $\lambda=2$  &   \hskip0.2cm 1.11  & 1.54\\
      \hline
        $\lambda=3$  &   \hskip0.2cm 1.13  &1.44   \\
      \hline
             $\lambda=4$  &   \hskip0.2cm 1.14  &1.37\\
      \hline
\end{tabular}
}
\parbox{42mm}{
fit parameters for HS
\begin{tabular}{|c |c|  c|}
 \hline
&  \hskip0cm  ${\Omega^*(2,2)}$  \hskip3cm &   $\sigma_\mathrm{e}/\sigma$  \\
      \hline
 $\lambda=1$  &   \hskip0.2cm 1.06 &   1.00\\
      \hline
        $\lambda=2$  &   \hskip0.2cm 1.25  & 1.02\\
      \hline
        $\lambda=3$  &   \hskip0.2cm 1.74 &1.14 \\
      \hline
             $\lambda=4$  &   \hskip0.2cm 2.21  &1.11\\
      \hline
\end{tabular}
}
\caption{The values for the fit parameters, ${\Omega^{*(2,2)}}$ and  $\mu/\mu_\mathrm{e}$ obtained from  fitting of the  simulation data to the DHS theory  (left)
as well as ${\Omega^{*(2,2)}}$ and $\sigma_\mathrm{e}/\sigma$ obtained from  fitting  the  simulation data to the Enskog theory for HS (right).}
\label{fit_eta}
\end{table}

The only previously available theory for viscosity of dense fluids is the HS Enskog theory.
In order to compare our theory to this, we fit the simulations data of viscosity to the Enskog theory for HS, Eq.~(\ref{eq:ENSHS}), with the HS RDF given in Eq.~(\ref{eq:CSRDF}).
The collision integral ${\Omega^{*(2,2)}}$ is equal to unity for HS, but this is incorrect for DHS.
When Enskog theory for HS is applied to real molecules this is usually taken into account by allowing ${\Omega^{*(2,2)}}$ to deviate from unity and using it as a fit parameter, along with the effective diameter $\sigma_\mathrm{e}$, and we do the same here.
The results are shown in  Fig.~\ref{DHS_eta}(b) as lines. 
The fit parameters ${\Omega^{*(2,2)}}$ and  $\sigma_\mathrm{e}/\sigma$ are given in Table.~\ref{fit_eta}. 
Fig.~\ref{DHS_eta} clearly shows that our theory successfully describes the viscosity of dense fluids of DHS and captures qualitative behaviour that is not captured by previous HS theoretical results.

In summary, we have developed a kinetic theory for the shear viscosity of dense fluids of dipolar hard spheres (DHS).
In our theory, we have included the long-range electrostatic interactions explicitly.
Our theory captures the main effects of the dipole-dipole interaction on the viscosity, which were missing from previous theories.
We see from our simulations that the differences between DHS and HS are mainly due to local structure.
At low densities the DHS viscosity is lower due to clustering of the particles.
At high densities, the DHS show orientational ordering, leading to stronger interaction and a higher viscosity. Both of these effects are captured by our theory. Our theory is in agreement with
simulation results for  packing fractions below about $0.35-0.4$.

While we have focused on the viscosity, the RDF at contact is the crucial ingredient for the collision rate and consequently the density-dependence of all non-equilibrium properties of fluids.
Our kinetic theory should therefore also provide for accurate descriptions of other transport properties, such as thermal conductivity and diffusion coefficient.
Moreover, the approaches currently in use in applications, for viscosity as well as other transport coefficients, are all based on the simple HS results, even for much more complicated molecules.
Besides the fundamental understanding of transport in polar fluids, our theory can thus also lead to significant improvements in the accuracy of calculations of transport properties in practical applications.

\section*{Acknowledgements}
The work has been supported by National Infrastructure for Computational Science in Norway
(UNINETT Sigma2) with computer timed for the Center for High Performance Computing (NN9573K and NN9572K). The authors acknowledge The Research Council of Norway for NFR project number 275507  and The Faculty of Engineering, Norwegian University of Science and Technology (NTNU), for financial support. FP acknowledges Prof. Ekaterina Elfimova and Prof. Philip  Camp for their advise and provided data.  

\label{ed}

\bibliography{ref}

\begin{thebibliography}{10}

\bibitem{Viswanath:07:0}
D.~S. Viswanath, T.~K. Ghosh, D.~H.~L. Prasad, N.~V.~K. Dutt, and K.~Y. Rani,
  {\em Viscosity of Liquids; Theory, Estimaation, Experiment, and Data}.
\newblock Springer, 2007.

\bibitem{stephan:79:0}
K.~Stephan and K.~Lucas, {\em Viscosity of Dense Fluids}.
\newblock New York: Plenum, 1979.

\bibitem{Ashurst:75:00}
W.~T. Ashurst and W.~G. Hoover {\em {\it Phys. Rev. A}}, vol.~11(2), 1975.

\bibitem{Daschakraborty:12:00}
S.~Daschakraborty and R.~Biswas {\em {\it J. Chem. Sci.}}, vol.~124,
  pp.~763--771, 2012.

\bibitem{Theiss:19:00}
M.~Theiss and J.~Gross {\em {\it J. Chem. Eng. Data}}, vol.~64, pp.~827--832,
  2019.

\bibitem{Astrid:12:00}
A.~S. de~Wijn, N.~Riesco, G.~Jackson, J.-P.~M. Trusler, and V.~Vesovic {\em
  {\it J. Chem. Phys.}}, vol.~136, p.~074514, 2012.

\bibitem{vesovic:12:00}
R.~Umla, N.~Riesco, and V.~Vesovic {\em {\it Fluid Ph. Equilib.}}, vol.~334,
  pp.~89--96, 2012.

\bibitem{Bayer:08:00}
I.~S. Bayer {\em {\it Lubricants}}, vol.~6, p.~30, 2018.

\bibitem{Das:03:00}
S.~Das, X.~Banquy, B.~Zappone, G.~W. Greene, G.~D. Jay, and J.~N. Israelachvili
  {\em {\it Biomacromolecules}}, vol.~14, pp.~1669--1677, 2013.

\bibitem{Madden:78:00}
W.~G. Madden, D.~D. Fitts, and W.~R. Smith {\em {\it Mol. Phys.}}, vol.~35,
  pp.~1017--1027, 1978.

\bibitem{Lee:89:0}
P.~H. Lee and B.~M. Ladanyi {\em {\it J. Chem. Phys.}}, vol.~91, p.~7063, 1989.

\bibitem{Kusalik:88:0}
P.~G. Kusalik and G.~N. Patey {\em {\it J. Chem. Phys.}}, vol.~88, p.~7715,
  1988.

\bibitem{patey:85:00}
P.~H. Fries and G.~N. Patey {\em {\it J. Chem. Phys.}}, vol.~82, p.~429, 1985.

\bibitem{Elfimova:16:00}
E.~D. Vtulkina and E.~A. Elfimova {\em {\it Fluid Ph. Equilib.}}, vol.~417,
  pp.~109--114, 2016.

\bibitem{Chapman:52:0}
S.~Chapman and T.~G. Cowling, {\em The Mathematicaltheory of non-uniform
  gases}.
\newblock Cambridge university press, 1952.

\bibitem{spherocylinders}
C.~Curtiss and C.~Muckenfuss {\em {\it J. Chem. Phys.}}, vol.~26, p.~1619,
  1957.

\bibitem{dhs-low-density}
R.~Olmsted and C.~Curtiss {\em {\it J. Chem. Phys.}}, vol.~55, p.~3276, 1971.

\bibitem{Astrid:08:00}
A.~S. de~Wijn, V.~Vesovic, G.~Jackson, and J.-P.~M. Trusler {\em {\it J. Chem.
  Phys.}}, vol.~128, p.~204901, 2008.

\bibitem{KUMARI:09:0}
A.~Kumari, P.~K. Sinha, M.~K. Sinha, and T.~K. Dey {\em {\it Int. J. Chem.
  Sci.:}}, vol.~7, p.~2477, 2009.

\bibitem{Castillo:93:00}
R.~Castillo and J.~Orozco vol.~79, pp.~344--357, 1993.

\bibitem{Santos:16:0}
A.~Santos, {\em A Concise Course on the Theory of Classical Liquids}.
\newblock Springer, 2016.

\bibitem{Sengers:00:0}
J.~V. Sengers, R.~F. Kayser, C.~J. Peters, and H.~J. White, {\em Equations of
  State for Fluids and Fluid Mixtures}.
\newblock Elsevier, 2000.

\bibitem{Lucas:79:0}
K.~Stephan and K.~Lucas, {\em Viscosity of Dense Fluids}.
\newblock Springer, 1979.

\bibitem{Carnahan:69:00}
N.~F. Carnahan and K.~E. Starling {\em {\it J. Chem. Phys.}}, vol.~51,
  p.~635–636, 1969.

\bibitem{monchick:61:0}
L.~Monchick and E.~A. Mason {\em {\it J. Chem. Phys.}}, vol.~35, p.~1676, 1961.

\bibitem{chung:84:00}
T.~Chung, L.~L. Lee, and K.~E. Starling {\em {\it Ind. Eng. Chem. Fundam}},
  vol.~23, pp.~8--13, 1984.

\bibitem{Ivanov:07:00}
A.~O. Ivanov and E.~V. Navok {\em {\it Colloid J.}}, vol.~69, p.~302, 2007.

\bibitem{Elfimova:10:1}
E.~A. Elfimova, A.~O. Ivanov, and C.~Holm {\em {\it J. Exp. Theor. Phys.}},
  vol.~111, pp.~146--156, 2010.

\bibitem{Elfimova:10:0}
J.~J. Cerda, E.~Elfimova, V.~Ballenegger, E.~Kruikova, A.~Ivanov, and C.~Holm
  {\em {\it Phys. Rev. E}}, vol.~81, p.~011501, 2010.

\bibitem{Elfimova:12:00}
E.~A. Elfimova and A.~O. Ivanov {\em {\it Phys. Rev. E}}, vol.~86, p.~021126,
  2012.

\bibitem{Ladanyi:99:00}
D.~V. Matyushov and B.~Ladanyi {\em {\it J. Chem. Phys.}}, vol.~10, p.~994,
  1999.

\bibitem{Henderson:11:0}
D.~Henderson {\em {\it Condens. Matter Phys}}, vol.~14, p.~33001, 2011.

\bibitem{Pippo:77:00}
R.~D. Pippo, J.~R. Dorfman, J.~Kestin, H.~E. Khalifa, and E.~A. Mason {\em {\it
  Physica A}}, vol.~86, pp.~205--223, 1977.

\bibitem{jover:12:0}
J.~Jover, A.~J. Haslam, A.~Galindo, G.~Jackson, and E.~A. M{\"u}ller {\em {\it
  J. Chem. Phys.}}, vol.~137, p.~144505, 2012.

\bibitem{vega:13:0}
J.~R. Espinosa, E.~Sanz, C.~Valeriani, and C.~Vega {\em {\it J. Chem. Phys.}},
  vol.~139, p.~144505, 2013.

\bibitem{faezeh:19:00}
F.~Pousaneh and A.~S. de~Wijn {\em {\it Mol. Phys.}}, vol.~118, p.~1622050,
  2020.

\bibitem{Drunsel:14:00}
F.~Drunsel, W.~Zmpitas, and J.~Gross {\em {\it J. Chem. Phys.}}, vol.~141,
  p.~054103, 2014.

\bibitem{Ballenegger:04:00}
V.~Ballenegger and J.-P. Hansen {\em {\it Mol. Phys.}}, vol.~102, pp.~599--609,
  2004.

\bibitem{Rushbrooke:73:00}
G.~S. Rushbrooke, G.~Stell, and J.~S. Høye vol.~26, p.~1199, 1973.

\bibitem{jog:99:00}
P.~K. Jog and W.~G. Chapman {\em {\it Mol. Phys.}}, vol.~97, pp.~307--319,
  1999.

\bibitem{palmer:94:0}
B.~J. Palmer {\em {\it Phys. Rev. E}}, vol.~49, p.~1, 1994.

\bibitem{Weis:93:00}
J.~J. Weis and D.~Levesque {\em {\it Phys. Rev. E}}, vol.~48 (5), 1993.

\bibitem{Weis:03:00}
J.~J. Weis {\em {\it J. Phys.:Cond. Mat}}, vol.~15, pp.~S1471--S1495, 2003.

\bibitem{Weis:06:00}
J.~J. Weis and D.~Levesque {\em {\it Nature Materials}}, vol.~125, p.~034504,
  2006.

\bibitem{Camp:00:00}
P.~J. Camp and G.~N. Patey {\em {\it Phys. Rev. E}}, vol.~62 (4), 2000.

\bibitem{wei:11:00}
D.~Wei, L.~Gao, J.~Zhang, L.~Yan, J.~Hu, L.~Chen, Z.~Gong, Y.~Guo, and Y.~Han
  {\em {\it Phys. Rev. E}}, vol.~83, p.~061703, 2011.

\bibitem{Klokkenburg:07:00}
M.~Klokkenburg, B.~H. Erne, A.~Wiedenmann, A.~V. Petukhov, and A.~P. Philipse
  {\em {\it Phys. Rev. E}}, vol.~75, p.~051408, 2007.

\bibitem{Klokkenburg:06:00}
M.~Klokkenburg, R.~P.~A. Dullens, W.~K. Kegel, B.~H. Erne, , and A.~P. Philipse
  {\em {\it Phys. Rev. Lett}}, vol.~96, p.~037203, 2006.

\bibitem{Butter:03:00}
K.~Butter, P.~H. Bomans, P.~M. Frederik, G.~J. Vroege1, and A.~P. Philipse {\em
  {\it J. Phys.:Cond. Mat}}, vol.~15, pp.~S1451--S1470, 2003.

\bibitem{Butter:03:01}
K.~Butter, P.~H. Bomans, P.~M. Frederik, G.~J. Vroege1, and A.~P. Philipse {\em
  {\it Nature Materials}}, vol.~2, pp.~88--91, 2003.

\end{thebibliography}
\bibliographystyle{ieeetr}

\end{document}